\documentclass[prl,aps,10pt,twocolumn,floatfix]{revtex4-1}
\usepackage{graphicx}
\begin{document}

\title{Quantum effects on one-dimensional collision dynamics of fermion clusters}

\author{Jun'ichi Ozaki}
\email{ozaki@scphys.kyoto-u.ac.jp}
\author{Masaki Tezuka}
\author{Norio Kawakami}

\affiliation{Department of Physics, Kyoto University, Kyoto 606-8502, Japan}

\begin{abstract}
Recently, many experiments with cold atomic gases have been conducted from interest in the non-equilibrium dynamics
of correlated quantum systems. Of these experiments, the mixing dynamics of fermion clusters motivates us to 
research cluster--cluster collision dynamics in one-dimensional Fermi systems.
We adopt the one-dimensional Fermi--Hubbard model and apply the time-dependent density matrix renormalization group method. 
We simulate collisions between two fermion clusters of spin-up and spin-down, and
calculate reflectance of the clusters $R$ changing the particle number in each cluster and the interaction strength between two fermions with up and down spins. 
We also evaluate the quasi-classical (independent collision) reflectance $R^\mathrm{qc}$ to compare it with $R$. 
The quasi-classical picture is quantitatively valid in the limit of weak interaction, but it is not valid when interaction is strong. 
\end{abstract}

\maketitle

\section{Introduction}

In recent years non-equilibrium dynamics of cold atom systems has emerged as interesting problems, because
the cold atom systems are ideally isolated systems which can be directly explored experimentally. 
Recent rapid progress in manipulation techniques has enabled experimentalists to modify strength and sign of interaction between cold atoms by Feshbach resonance \cite{feshbach resonance}, 
to make a spin-dependent trapping potential to trap the cold atoms of each spin separately,
and to change the trapping potential suddenly to explore the dynamics of quantum quench. 

Of these experiments, the dynamics of the mixing of two spin components of fermion gas \cite{spin-mixing exp} motivates us 
to study collision dynamics between two fermionic atom clusters in one-dimensional Fermi systems. 
The fermion cluster--cluster collision dynamics is so simple to compare it with classical dynamics, 
but the quantum effects such as Fermi statistics on collision dynamics have not been clarified. 
Furthermore, cluster collision dynamics is one of the basic concepts of dynamics, which can be applied to diverse quantum dynamics. 
Therefore the research of fermion cluster collision dynamics brings a new perspective of quantum dynamics.

\begin{figure}[htbp]
\begin{center}
\includegraphics[width=8cm,clip]{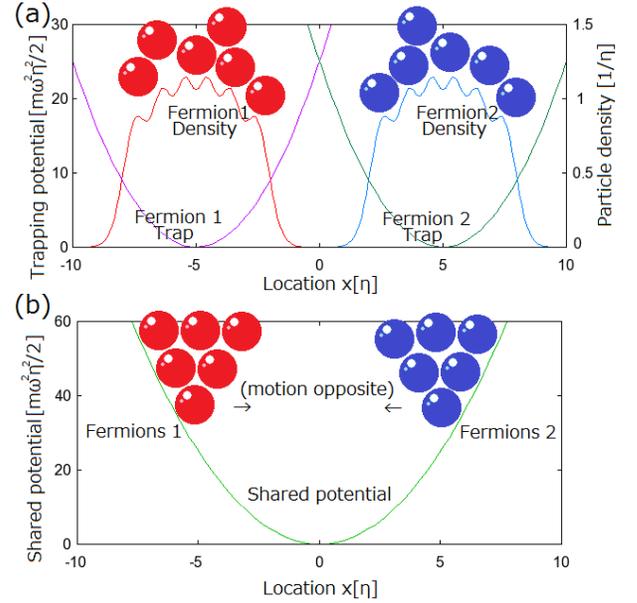}
\caption{
(a) Trapping potentials of the two types of fermions and particle density of the trapped fermion clusters ($n=6$), and (b) the shared potential. 
}
\label{trap}
\end{center}
\end{figure}

\section{Theoretical conditions}
Simulations are conducted in the one-dimensional system which consists of two-component fermions. 
In the system, at first, the fermions of each component are trapped by separate harmonic potentials. 
The trapping harmonic potentials have the same shape (the cycle of motion of a fermion in the potential is $T$), but they are spatially separated (Fig.\ref{trap}(a)).
The particle density has some peaks (the number of peaks is the same as the particle number) and has Gaussian-like tails. 

\begin{figure}[htbp]
\begin{center}
\includegraphics[width=8cm,clip]{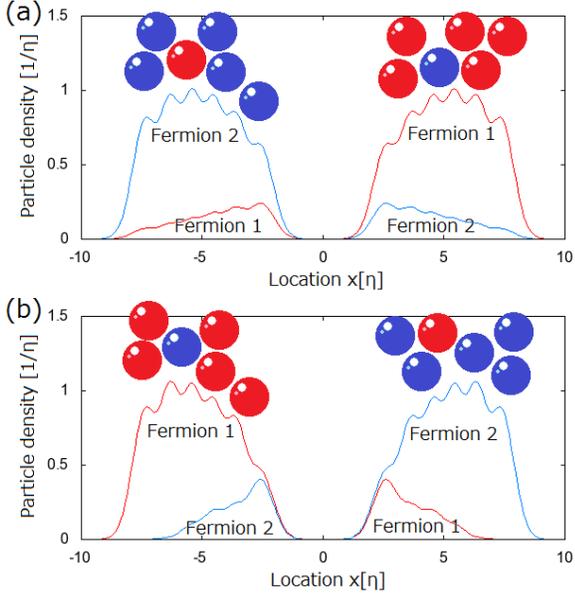}
\caption{
(a) Particle density at $t=T/2$ and the interaction $u=9.05 [\hbar \eta T^{-1}]$ ($n=6$), and
(b) particle density at $t=T/2$ and interaction $u=144.8 [\hbar \eta T^{-1}]$ ($n=6$). 
}
\label{action}
\end{center}
\end{figure}

Then, we suddenly change the trapping potentials into a new shared potential (we set this time $t=0$). 
The new shared potential has the same shape as before, but its bottom is midway between the bottoms of the two initial trapping potentials (Fig.\ref{trap}(b)). 
When the interaction between the two fermions is zero, each fermion cluster moves to the opposite side in the half cycle $T/2$, without changing the shape of the particle number density. 
On the other hand, if the interaction is finite, some particles are reflected and return to the initial side, and the others pass through and move to the opposite side (Fig.\ref{action}(a)(b)). 
At $t=T/2$, the particle number density is nearly zero again at the center of the potential, so we can obtain the number of particles which return to the initial side; 
this number is denoted by $N_n^\mathrm{ref}$ ($n$ is the number of particles in the initial cluster). Therefore we can obtain the reflectance $R_n \equiv N_n^\mathrm{ref} / n$, and the transmittance $T_n \equiv 1-R_n$. 
Clearly, if the interaction between the two fermions is zero, $R_n = 0$, and if the interaction is infinite, $R_n = 1$ because the system is one-dimensional.

\section{Simulations by t-DMRG method}
We adopt the one-dimensional Fermi--Hubbard model and apply the time-dependent density matrix renormalization group(t-DMRG) method to simulate the system. 
We take $N_\mathrm{site} \equiv 4D+1$ sites numbered $0, 1, ... , 2D, ... , 4D$ ($D$ is an integer), and the system size is $L=4a$; 
the site 0 is at $x=-2a$, the site $2D$ is at $x=0$, and the site $4D$ is at $x=2a$. 
The lattice constant is $\Delta x \equiv a/D$. 
The trapping potential is $V_-^\mathrm{ext}(x) \equiv \frac{1}{2}m\omega^2(x+a)^2$ for the spin-down particles, and
$V_+^\mathrm{ext}(x) \equiv \frac{1}{2}m\omega^2(x-a)^2$ for the spin-up particles, where $m$ is the particle mass, and $\omega \equiv 2\pi/T$ and $T$ is the cycle of the oscillation. 
The continuum limit is recovered in the limit of $N_\mathrm{site} \rightarrow \infty$. 
The edge of the particle density decays in about $\eta \equiv 1/\sqrt{m\omega}$. 
At $t=0$, we change the trapping potentials to $V_\pm^\mathrm{ext}(x) \equiv \frac{1}{2}m\omega^2x^2$. 
The Hamiltonian is
\begin{eqnarray*}
H &=& -J\sum_{i,\sigma} (a^\dagger_{\sigma,i}a_{\sigma,i+1} + a^\dagger_{\sigma,i+1}a_{\sigma,i})\\ &+& U \sum_{i} n_{+,i} n_{-,i} + \sum_{i,\sigma}V_\sigma^\mathrm{ext}(x_i)n_{\sigma,i} \;.
\end{eqnarray*}
Here $J \equiv (2m\Delta x^2)^{-1}$ is the hopping constant, and $U$ is the on-site interaction between a spin-up fermion and a spin-down fermion. 
In the continuum limit, the interaction term reproduces the contact interaction $u \delta(x_1-x_2)$, in which $u \equiv U\Delta x$ is the interaction strength.

Here we show the results for the system size $L = 20\eta$ ($a=5\eta$) and $D=50$. 
The system size is large enough compared to the clusters' expanse, so the particles do not escape from the system, 
and we assume that $D$ is large enough so that the effect of the discretization is insignificant.

\section{Results} 

\begin{figure}[htbp]
\begin{center}
\includegraphics[width=8cm,clip]{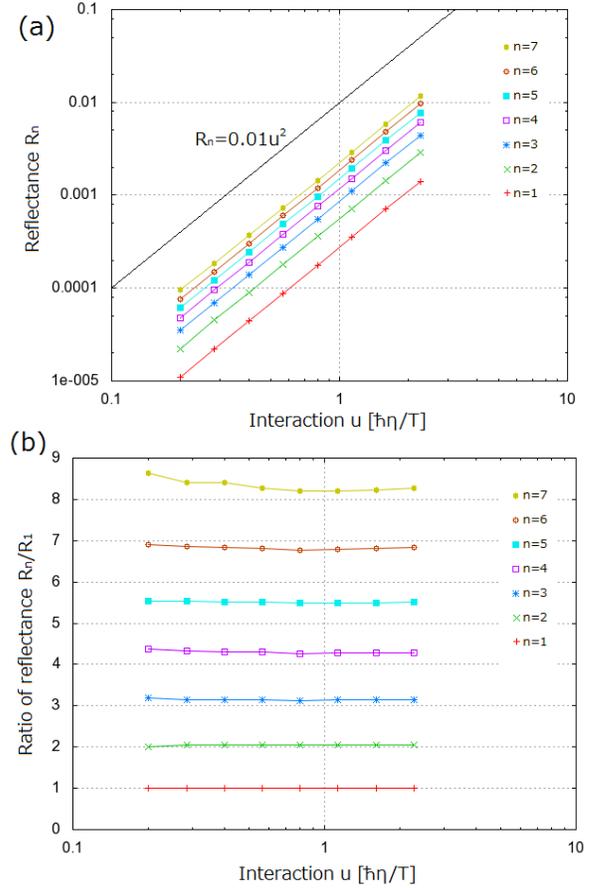}
\caption{
(a) Reflectance $R_n$ for $u = U\Delta x$ between 0.200 and 2.260 $[\hbar \eta T^{-1}]$, and (b) ratio of reflectance $R_n/R_1$. 
}
\label{result_small}
\end{center}
\end{figure}

\begin{figure}
\begin{center}
\includegraphics[width=8cm,clip]{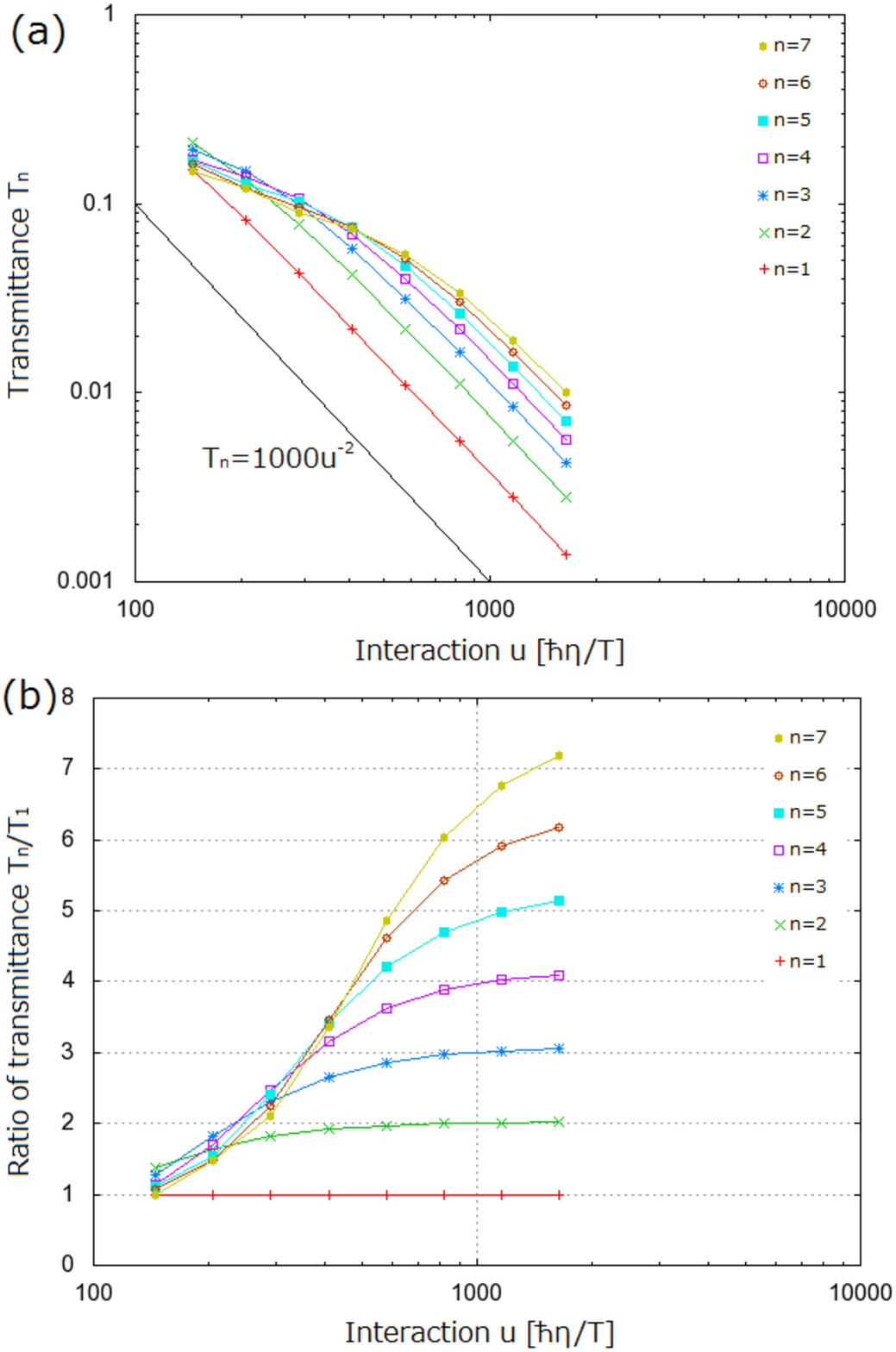}
\caption{
(a) Transmittance $T_n$ for $u = U\Delta x$ between 144.8 and 3276.8 $[\hbar \eta T^{-1}]$, and (b) ratio of transmittance $T_n/T_1$. 
}
\label{result_large}
\end{center}
\end{figure}

Figure \ref{result_small}(a) shows the reflectance $R_n$ for the interaction $u$ between 0.200 and 2.260 $[\hbar \eta T^{-1}]$. 
The simple theoretical calculation predicts $R_1 \propto u^2$. The figure implies $R_n \propto u^2$ for any $n$. 
We plot the ratio of $R_n$ to $R_1$ in Fig.\ref{result_small}(b), and it is observed that $R_n \simeq n R_1$ in the limit of $u \rightarrow 0$. 
Therefore the results indicate $R_n \propto nu^2$ as $u \rightarrow 0$. 

On the other hand, Fig.\ref{result_large}(a) shows the transmittance $T_n$ for $u$ between 144.8 and 3276.8 $[\hbar \eta T^{-1}]$. 
Theoretically, $T_1 \propto u^{-2}$. The figure indicates $T_n \propto u^{-2}$ as $u \rightarrow \infty$ for all $n$. 
Figure \ref{result_large}(b) displays $T_n / T_1$, and we find the relation $T_n \simeq n T_1$ as $u \rightarrow \infty$. 
We observe that $T_n \propto nu^{-2}$ in the limit of strong on-site interaction. 

The numerical results indicate that the sign of $u$ has no effect within the computational accuracy. 
We also calculate for $D=25$, and the results $R_n$ and $T_n$ are changed typically by 10\%. 
For smaller values of $D$, the wave function loses its original form faster as time advances because of the effect of the discretization. 
However the relations $R_n = n R_1$ and $T_n = n T_1$ are still observed.

\section{Discussion} 
We assume a simple quasi-classical dynamics model to compare it with the calculation results. 
The assumption of the quasi-classical model is that the fermions in the clusters are independently localized in each position, 
and the collision dynamics is described by a series of one-to-one collisions between two wave packets of the independent fermions. 
Under this assumption of ``quasi-classical independent collision'', we can calculate quasi-classical reflectance $R_n^\mathrm{qc}$ 
only from the one-to-one collision reflectance $R_1$ (by definition $R_1^\mathrm{qc} = R_1$). 

We observe from this simple calculation that $R_n^\mathrm{qc} \propto nu^2$ in the limit of $u \rightarrow 0$. 
The reason is that a wave packet of a fermion collides $n$ times with fermions of the other spin, 
while it moves to the opposite side (note that almost all probability amplitude of it reaches the opposite). 
Therefore the total number of collisions is $n^2$ so the reflectance $R_n^\mathrm{qc} = (n^2R_1)/n = nR_1$. 
This result is consistent with our simulation results by t-DMRG. 

In contrast, in the limit of $u \rightarrow \infty$, it is shown that $1-R_n^\mathrm{qc} \equiv T_n^\mathrm{qc} \propto u^{-2}$ and $T_n^\mathrm{qc}$ is independent of $n$. 
In this limit, the particles are almost perfectly reflected by each other, so passing to the opposite side occurs only on the border between the two clusters. 
Since the collision on the surface occurs just $n$ times, the transmittance $T_n^\mathrm{qc} = (nT_1)/n = T_1$. 
This result conflicts with our simulation with the quantum effects fully considered. 

Therefore the quasi-classical picture, ``quasi-classical independent collision'', is not valid in the limit of $u \rightarrow \infty$, 
so a different picture to explain the results for $u \rightarrow \infty$ is needed. 
Since the particles are fermions, the maximum number of particles which share a site is just two. 
These results imply that the number of possible one-to-one collisions on the surface, which is $n^2$, determines both 
the reflectance in the limit of $u \rightarrow 0$ and the transmittance in the limit of $u \rightarrow \infty$ 
because of dynamically emerging quantum many-body effects.

\section{Conclusion} 
Using the time-dependent density matrix renormalization group method and the Fermi--Hubbard model, 
we calculate collision dynamics between two kinds of fermion clusters in both cases of weak and strong interaction. 
Reflectance (transmittance) is in proportion to the particle number when the interaction is weak (strong).
At least in the two extremes, the cluster--cluster collision dynamics is mainly dependent on the number of possible one-to-one collisions. 

This work is partly supported by KAKENHI (Nos. 21740232, 20104010) and JSPS through its FIRST Program.

\end{document}